\documentclass[sigconf]{acmart}

\AtBeginDocument{%
  \providecommand\BibTeX{{%
    \normalfont B\kern-0.5em{\scshape i\kern-0.25em b}\kern-0.8em\TeX}}}

\copyrightyear{2021} 
\acmYear{2021} 
\setcopyright{acmcopyright}\acmConference[CIKM '21]{Proceedings of the 30th ACM International Conference on Information and Knowledge Management}{November 1--5, 2021}{Virtual Event, QLD, Australia}
\acmBooktitle{Proceedings of the 30th ACM International Conference on Information and Knowledge Management (CIKM '21), November 1--5, 2021, Virtual Event, QLD, Australia}
\acmPrice{15.00}
\acmDOI{10.1145/3459637.3482242}
\acmISBN{978-1-4503-8446-9/21/11}

\usepackage{xspace}
\usepackage{amsthm}
\usepackage{amsmath}
\usepackage{array}
\usepackage{bm}
\newtheorem{definition}{Definition}[section]

\newcommand{\modelname}{\textsf{TGSRec}\xspace}
\newcommand{\layername}{\textsf{TCT}\xspace}
\usepackage{tabulary}
\usepackage{graphicx}
\usepackage{soul}
\usepackage[normalem]{ulem}
\useunder{\uline}{\ul}{}
\usepackage{enumitem}
\usepackage{lipsum}
\usepackage{booktabs}
\usepackage{caption}
\usepackage{subcaption}
\usepackage{multirow}
\usepackage{balance}
\usepackage[flushleft]{threeparttable}

\setlist[itemize]{leftmargin=*}
\graphicspath{ {../fig/} }

\begin{document}
\fancyhead{}
\title{Continuous-Time Sequential Recommendation with Temporal Graph Collaborative Transformer}

\author{Ziwei Fan*, Zhiwei Liu*}\thanks{*Both authors contribute equally.}
\affiliation{%
  \institution{Department of Computer Science, University of Illinois at Chicago}
  \country{USA}
}
\email{{zfan20,zliu213}@uic.edu}

\author{Jiawei Zhang}
\affiliation{%
  \institution{IFM Lab, Department of Computer Science, University of California, Davis
}
    \country{USA}
}
\email{jiawei@ifmlab.org}

\author{Yun Xiong}
\affiliation{%
  \institution{Shanghai Key Laboratory of Data Science, School of Computer Science, 
Fudan University}
    \country{China}
}
\email{yunx@fudan.edu.cn}

\author{Lei Zheng}
\affiliation{%
  \institution{Pinterest Inc.}
    \country{USA}
}
\email{lzheng@pinterest.com}

\author{Philip S. Yu}
\affiliation{%
  \institution{Department of Computer Science, University of Illinois at Chicago}
  \country{USA}
}
\email{psyu@uic.edu}





\begin{abstract}

In order to model the evolution of user preference, we should learn user/item embeddings based on time-ordered item purchasing sequences, which is defined as Sequential Recommendation~(SR) problem.
Existing methods leverage sequential patterns to model item transitions. 
However, most of them ignore crucial temporal collaborative signals, 
which are latent in evolving user-item interactions and coexist with sequential patterns. 
Therefore, we propose to unify sequential patterns and temporal collaborative signals to improve the quality of recommendation, which is rather challenging.
Firstly, it is hard to simultaneously encode sequential patterns and collaborative signals. Secondly, it is non-trivial to express the temporal effects of collaborative signals. 

Hence, we design a new framework \textbf{T}emporal \textbf{G}raph \textbf{S}equential \textbf{Rec}ommender~(\modelname) upon our defined continuous-time bipartite graph. We propose a novel Temporal Collaborative Transformer~(\layername) layer in \modelname, which advances the self-attention mechanism by adopting a novel collaborative attention. \layername layer can simultaneously capture collaborative signals from both users and items, as well as considering temporal dynamics inside sequential patterns.
We propagate the information learned from \layername layer over the temporal graph to unify sequential patterns and temporal collaborative signals. Empirical results on five datasets show that \modelname significantly outperforms other baselines, in average up to $22.5\%$ and $22.1\%$ absolute improvements in Recall@$10$ and MRR, respectively. Our code is available online in \url{https://github.com/DyGRec/TGSRec}.
\end{abstract}



\maketitle

\section{Introduction}
Recommender system has become essential in providing personalized information filtering services in a variety of applications~\cite{liang2016modeling,wang19neural,Liu2019JSCNJS, wang2021dkg, peng2020m2}. It learns the user and item embeddings from historical records on the user-item interactions~\cite{he2017neural,rendle2009bpr}. In order to model the dynamics of the user-item interaction, current research works~\cite{rendle2010factorizing,wu2017recurrent,hidasi2015session,tang2018personalized, fan2021Modeling} leverage historical time-ordered item purchasing sequences to predict future items for users, referred to as the sequential recommendation~(SR) problem~\cite{hidasi2015session,kang2018self}. One of the fundamental assumptions of SR is that the users' interests change smoothly~\cite{hidasi2015session,wu2017recurrent,tang2018personalized,kang2018self}. Thus, we can train a model to infer the items more likely to appear in the future sequence. For example, with the recent developments of \textit{Transformer}~\cite{vaswani2017attention}, current endeavors design a series of self-attention SR models to predict future item sequences~\cite{kang2018self,sun2019bert4rec,ssept20wu}. A self-attention model infers sequence embeddings at position $t$ by assigning an attention weight to each historical item and aggregating these items. The attention weights reveal impacts of previous items to the current state at time point $t$. 

\begin{figure}
    \centering
    \includegraphics[width=0.35\textwidth]{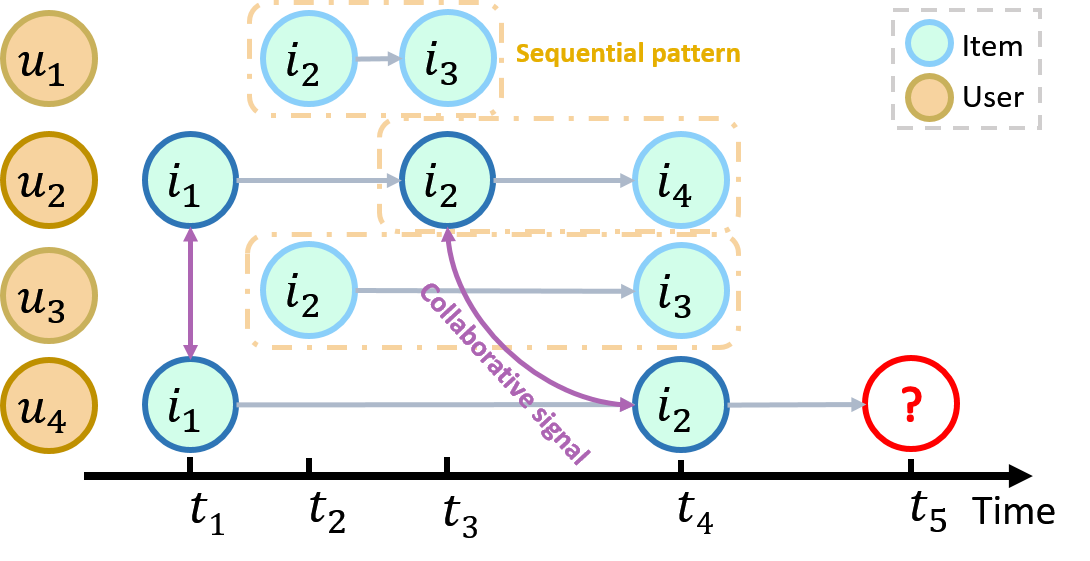}
    \caption{A toy example of temporal collaborative signals. 
    Given the items that users $u_1, u_2, u_3$ and $u_4$ like in the past timestamps $t_1, t_2, t_3$ and $t_4$, the target is to recommend an item to $u_4$ at $t_5$ as the next item after $i_2$. 
    }
    \label{fig:toy_exp}
\end{figure}

Despite their effectiveness, existing works only leverage the sequential patterns to model the item transitions within sequences, which is still insufficient to yield satisfactory results. The reason is that they ignore the crucial \textbf{temporal collaborative signals}, which are latent in evolving user-item interactions and coexist with sequential patterns. To be specific, we present the effects of temporal collaborative signals in Figure~\ref{fig:toy_exp}. The target is to recommend an item to $u_4$ at $t_5$ as the next item after $i_2$. By only considering the sequential patterns, $i_3$ is recommended 
since it appears $2$ times after $i_2$ as in $u_1$ and $u_3$, compared with $i_4$ of only $1$ time in $u_2$. However, if also taking account of collaborative signals, we would recommend $i_4$, because both $u_2$ and $u_4$ have interactions with $i_1$ at $t_2$ and $i_2$ at $t_3$ and $t_4$, respectively, which indicates their high similarity. Hence, $u_2$'s sequential patterns are of more impacts to $u_4$. This motivates us to \textit{unify sequential patterns and temporal collaborative signals}. 

However, incorporating temporal collaborative signals in SR is rather challenging. The first challenge is that it is hard to simultaneously encode collaborative signals and sequential patterns. Current models capture the sequential pattern based on the transition of items \text{within sequences}~\cite{hidasi2015session,kang2018self,ma2020disentangled}, thus lacking the mechanism to model the collaborative signals \text{across sequences}. Jodie~\cite{kumar2019predicting} and DGCF~\cite{li2020dynamic} employs LSTM to model the dynamics and interactions of user and item embeddings but they cannot learn the impacts of all historical items, thus unable to encode sequences. SASRec~\cite{kang2018self} proposed to use a self-attention mechanism to encode item sequence, while the effects of users, i.e., collaborative signals, are omitted. SSE-PT~\cite{ssept20wu} implicitly models collaborative signals by directly adding the same user embedding into the sequence encoding. However, it fails to model the interactions between user and item, thus unable to explicitly express collaborative signals.  

The second challenge is that it is hard to express the temporal effects of collaborative signals. In other words, it remains unclear how to measure the impacts of those signals from a temporal perspective. For example, in Figure~\ref{fig:toy_exp}, $i_1$ is interacted with $u_2$ and $u_4$ at $t_1$, while $i_2$ is interacted with them respectively at $t_3$ and $t_4$. Since there is a lag, it is problematic to ignore the time gap and assume they are of equal contributions. We should use temporal information to infer the importance of those collaborative signals to the recommendation on $t_5$. Existing works~\cite{hidasi2015session,kang2018self,sun2019bert4rec,li2020dynamic} assume that items appear discretely with equal time intervals. Thus, they only focus on the orders/positions of items in the sequence, which limits their capacity in expressing the temporal information. Some recent works~\cite{li2020time,ye2020time} also notice the importance of time span. But their models either fail to capture time differences between historical interactions or are unable to generalize to any unseen future timestamps or time difference, thus are still far from revealing the actual temporal effects of collaborative signals.



\begin{figure}
\begin{subfigure}[t]{0.22\textwidth}
    \
    \includegraphics[width=0.9\textwidth]{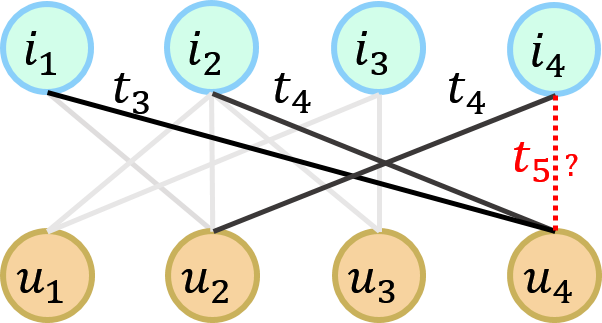}
    \caption{Example of CTBG}
    \label{fig:CTBG}
\end{subfigure}
\begin{subfigure}[t]{.22\textwidth}
    \includegraphics[width=1\textwidth]{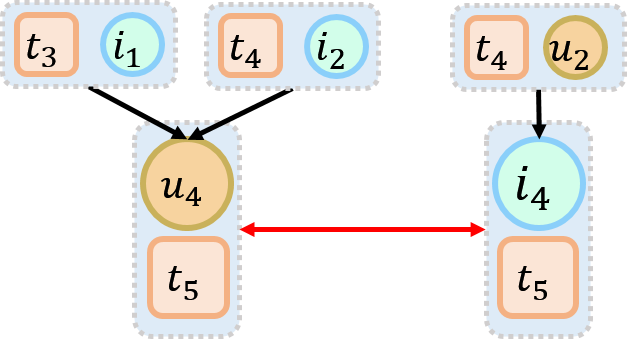}
    \caption{Temporal Inference}
    \label{fig:aggregation}
\end{subfigure}
\vspace{-3mm}
\caption{The associated CTBG of Figure~\ref{fig:toy_exp} and the inference of temporal embeddings of $u_4$ and $i_4$ at $t_5$.}
\end{figure}


Current transformer-based models~\cite{kang2018self,sun2019bert4rec} adopt self-attention mechanism, which has query, key, and value inputs from item embeddings and employs dot-product to learn their correlation scores. The limitation is that self-attention is only able to capture item-item relationships in sequences.
Additionally, they have no module to capture temporal correlations of items. 
To this end, we propose a new model \textbf{T}emporal \textbf{G}raph \textbf{S}equential \textbf{Rec}ommender~(\modelname). It consists of two novel components: (1) the Temporal Collaborative Transformer~(\layername) layer and (2) graph information propagation. 

The first component advances current transformer-based models as it can explicitly model collaborative signals in sequences and express temporal correlations of items in sequences. 
To be more specific, \layername layer adopts \textit{collaborative attention} among user-item interactions, where the query input to the collaborative attention is from the target node~(user\slash item), while the key and value inputs are from connected neighbors. As such, \layername layer learns the importance of those interactions,
thus well characterizing the collaborative signals. 
Moreover, \layername layer fuses the temporal information into the collaborative attention mechanism, 
which explicitly expresses the temporal effects of those interactions. Altogether, the \layername layer captures temporal collaborative signals. 

The second module is devised upon our proposed Continuous-Time Bipartite Graph (CTBG). The CTBG consists of user\slash item nodes, and interaction edges with timestamps, as shown in Figure~\ref{fig:CTBG}. Given timestamps, neighbor items of users preserve sequential patterns. We propagate temporal collaborative information learned around each node to surrounding neighbors over CTBG. Therefore, it unifies sequential patterns with temporal collaborative signals.


In this work, we propose to use temporal embeddings of nodes for recommendation, 
which are dynamic and inferred at specified timestamps. 
For example, at time $t$, we infer the temporal user embedding by aggregating the context. 
We illustrate the temporal inference of $u_4$ and $i_4$ at time $t_5$ in Figure~\ref{fig:aggregation}. 
The temporal embeddings are inferred by our proposed \layername layer.
It uses temporal information to discriminate impacts of those historical interactions and makes inferences of temporal node embeddings.
The contributions of this paper are as follows:\\
    \noindent \textbf{Graph Sequential Recommendation:} We connect the SR problem with graph embedding methods, which focuses on unifying the sequential patterns and temporal collaborative signals.\\
    \noindent \textbf{Temporal Collaborative Transformer:} We propose a novel temporal collaborative attention mechanism to infer temporal node embeddings, which jointly models collaborative signals and temporal effects. This overcomes the inadequacy of the traditional self-attention mechanism on capturing the temporal effects and user-item collaborative signals.\\
    \noindent \textbf{Extensive Experiments:} We conduct a comparison experiment on five real-world datasets. 
    Comprehensive experiments demonstrate the state-of-the-art performance of \modelname and its effectiveness of modeling temporal collaborative signals.
    

\section{Related Work}\label{sec:related_work}
In this section, we first review some related work, which includes sequential recommendation~(SR), temporal information, and some graph-based recommender systems. 

\subsection{Sequential Recommendation}
\label{subsec:sr}
SR predicts the future items in the user shopping sequence by mining the sequential patterns. An initial solution to the SR problem is to build a Recurrent Neural Network~(RNN)~\cite{hidasi2015session,wu2017recurrent,yu2016dynamic, 10.1145/3331184.3331329}. GRU4Rec~\cite{hidasi2015session} is proposed to predict the next item in a session by employing the GRU modules. Later, a Hierarchical RNN~\cite{quadrana2017personalizing} is proposed to enhance the RNN model regarding the personalizing information. Additionally, LSTM~\cite{hochreiter1997long,wu2017recurrent} can be applied to explore both the long-term and short-term sequential patterns. Moreover, in order to capture the intent of users at local sub-sequence, NARM~\cite{li2017neural} is proposed by combining the RNN model with attention weights. The major drawback of the RNN model is that it can only generate a single hidden vector, which limits its power to encode sequences~\cite{chen2018sequential}. 


Recently, owing to the success of self-attention model~\cite{vaswani2017attention,devlin2018bert, liu2021enriching, zhang2021pretrained} in NLP tasks, a series of attention-based SR models are proposed~\cite{kang2018self,sun2019bert4rec,ma2020disentangled,wu2020deja,ji2020hybrid, liu2021augmenting, peng2021ham}. SASRec~\cite{kang2018self} applies the transformer layer to assign weights to items in the sequence. Later, inspired by the BERT~\cite{devlin2018bert} model, BERT4Rec~\cite{sun2019bert4rec} is proposed with a bidirectional transformer layer. \cite{ma2020disentangled} introduce the sequence to sequence training procedure in SR. SSE-PT~\cite{ssept20wu} designs a personalized transformer to improve the SR performance. ASReP~\cite{liu2021augmenting} proposes augmenting short sequences to alleviate the cold-start issue in Transformer.  TiSASRec~\cite{li2020time} enhances SASRec with the time-interval information found in the training data. However, these models only focus on the item transitions within sequences, while unable to unify the important temporal collaborative signals with sequential patterns and are not generalized to unseen timestamps. 

\subsection{Temporal Information}
Previously mentioned SR works are specifically designed to capture sequential patterns, while ignoring the important temporal information~\cite{koren2009collaborative,xiong2010temporal,xiang2010temporal,kumar2019predicting,li2020time}. In practice, the context of users and items changes over time, which is crucial for modeling the temporal dynamics in SR. TimeSVD++~\cite{koren2009collaborative} is a representative work which models the temporal information into collaborative filtering~(CF) method. It simply treats the bias as a function over time. BPTF~\cite{xiong2010temporal} extends matrix factorization to tensor factorization and uses time as the third dimension. MS-IPF~\cite{xiang2010temporal} defines a temporal graph, where it operates PageRank algorithm for recommendation. Recently, CDTNE~\cite{nguyen2018continuous} is proposed by applying temporal random walk over its defined continuous-time dynamic network. TGAT~\cite{xu2020inductive} also introduces temporal attention for learning dynamic graph embeddings. 
JODIE~\cite{kumar2019predicting} develops user and item RNNs to update user and item embeddings. Regarding the SR problem, a few recent works~\cite{li2020time,ye2020time} also notice the importance of temporal information. CTA~\cite{wu2020deja}, MTAM~\cite{ji2020hybrid}, and TiSASRec~\cite{li2020time} all consider to use time intervals between successive items in sequences.  TASER~\cite{ye2020time} encodes both the absolute time and relative time as vectors, which are processed to attention models to complete the SR task. However, these models are not able to unify temporal collaborative signals with sequential patterns.

\subsection{Graph-based Recommendation}
Because we solve the SR problem based on the graph structure~\cite{zhang2017learning,zhang2019ige}, we also review some graph-based recommender system models~\cite{nguyen2018continuous,wang19neural,he2020lightgcn,Liu2019JSCNJS,liu2020basconv, 9377917}, especially those based on Graph Neural Network~(GNN) methods~\cite{kipf17semi,wang19neural,he2020lightgcn,berg2017gcmc}. 
Compared with directly learning from sequences, graph-based models can also capture the structural information~\cite{nguyen2018continuous,berg2017gcmc}. Both NGCF~\cite{wang19neural} and LightGCN~\cite{he2020lightgcn} argue that graph-based models are able to effectively model collaborative signals, which is crucial in learning user/item embeddings. The successes of GNN in recommender systems~\cite{wang2019kgat,he2020lightgcn,wang19neural,berg2017gcmc} provide simple yet effective methods in learning user/item embeddings from graphs. GNN models learn the embeddings by aggregating neighbors~\cite{wang19neural,he2020lightgcn}. Therefore, it is easy to stack multiple layers to learn both the first-order and high-order collaborative signals~\cite{wang19neural,he2020lightgcn}. CTDNE~\cite{nguyen2018continuous} defines a temporal graph to learn dynamic embeddings of nodes. TGAT~\cite{xu2020inductive} learns the dynamic graph embeddings based on the graph attention model. Basconv~\cite{liu2020basconv} characterizes heterogeneous graphs to learn user/item embeddings. Those models argue that graph is powerful in modeling both the structural and temporal information. However, few works investigate the possibility of solving SR problems based on graphs. SR-GNN~\cite{wu2019session} learns embeddings of session graphs by using a GNN to aggregate item embeddings but fails to model temporal collaborative signals.   

\section{Definitions and Preliminaries}
In this section, we introduce some definitions and preliminaries. Different from using users' interactions sequences as inputs in SR, we introduce the \textit{Continuous-Time Bipartite Graph}~(CTBG) to represent all temporal interactions. Each edge in this graph has the timestamp as the attribute. The directly connected neighbors of every user\slash item node in this graph preserve the sequential order via the timestamps at edges. The formal definition of CTBG are given in the following:

\begin{definition}[Continuous-Time Bipartite Graph]
A continuous time bipartite graph with $N$ nodes and $E$ edges for recommendations is defined as $\mathcal{B} = \{\mathcal{U}, \mathcal{I}, \mathcal{E_T}\}$, where $\mathcal{U}$ and $\mathcal{I}$ are two disjoint node sets of users and items, respectively. Every edge $e\in \mathcal{E_T}$ is denoted as a tuple $e=(u,i,t)$, where $u\in\mathcal{U}$, $i\in\mathcal{I}$, and $t\in \mathbb{R}^+$ as the edge attribute. Each triplet $(u,i,t)$ denotes the interaction of a user $u$ with item $i$ at timestamp $t$.
\end{definition}

This paper focuses on the SR problem with continuous timestamps. An example of the CTBG is presented in Figure~\ref{fig:CTBG}. Let $\mathcal{I}_u(t)$ denote the set of items interacted with the user $u$ before timestamp $t$, and $\mathcal{I} \setminus \mathcal{I}_u(t)$ denote the remaining items. We defined the continuous-time sequential recommendation problem which we study in this paper as following:
\begin{definition}[Continuous-Time Recommendation]
At a specific timestamp $t$, given user set $\mathcal{U}$, item set $\mathcal{I}$, and the associated CTBG, the continuous-time recommendation of $u$ is to generate a ranking list of items from $\mathcal{I} \setminus \mathcal{I}_u(t)$, where the items that $u$ is interested will be ranked top in the list.
\end{definition}

Then, the SR problem is equivalent to make continuous-time recommendations on a set of future timestamps $\mathcal{T}_{u}$ for each user $u$:
\begin{definition}[Continuous-Time Sequential Recommendation]\label{def:ct_seq_rec}
For a specific user $u$, given a set of future timestamps $\mathcal{T}_{u} > T$, the continuous-time sequential recommendation for this user is to make a continuous-time recommendation for every timestamp $t\in \mathcal{T}_{u}$.
\end{definition}

This is a generalized definition compared with other works~\cite{kang2018self,ma2020disentangled}. We explicitly consider timestamps, while others only care about the orders/positions. Therefore, differing from existing works using next-item prediction to evaluate sequential recommendation, future timestamps should be present to make a prediction.  If timestamps are position numbers in sequences, the studied problem is reduced to the same definition as using only orders/positions information. Note that timestamp can be any real value, thus being continuous. 

\section{Proposed Model}

\begin{figure*}[htp]
    \centering
    \includegraphics[width=0.9\textwidth]{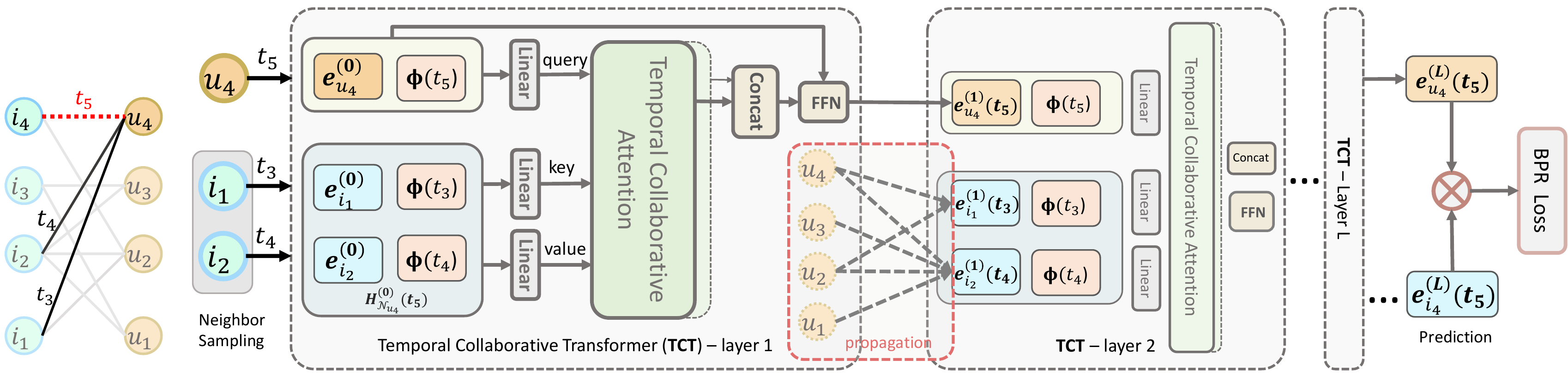}
    \vspace{-3mm}
    \caption{The framework of \modelname. The query node is $u_4$, whose final temporal embedding at time $t_5$ is $\bm{h}_{u_4}^{(2)}(t_5)$. The \layername layer samples its neighbor nodes and edges. Timestamps on edges are encoded as vectors by using mapping function $\Phi$. Node embeddings for the first \layername layer are long-term embeddings. Node embeddings for other \layername layers (\textit{e.g.} layer 2) are propagated from the previous \layername layer, thus being temporal node embeddings.
    }
    \label{fig:architect_agg}
\end{figure*}

In this section, we present the \modelname model, which unifies sequential patterns and temporal collaborative signals. The framework of the \modelname model is presented in Figure~\ref{fig:architect_agg}. There are three major components: 1) Embedding layer, which encodes nodes and timestamps in a consistent way to connect the SR problem with graph embedding method; 2) Temporal Collaborative Transformer~(\layername) layer, which employs a novel temporal collaborative attention mechanism to discriminate temporal impacts of neighbors, and aggregates both node and time embeddings to infer the temporal node embedding; 3) Prediction layer, which utilizes output embeddings from the final \layername layer to calculate the score. 

\subsection{Embedding Layer}
We encode two types of embeddings in this paper, one being the \textit{long-term embeddings} of nodes, and the other being the \textit{continuous-time embeddings} of timestamps on edges. %

\subsubsection{Long-Term User\slash Item Embeddings}
Long-term embeddings for users and items are necessary~\cite{devooght2017long} for long-term collaborative signals representation. 
In CTBG, it functions as node features and is optimized to model the holistic structural information. A user (item) node  is parameterized by a vector $\bm{e}_u(\bm{e}_i)\in \mathbb{R}^d$. Since we learn embeddings for nodes in the CTBG, we retrieve the embedding of a node by indexing an embedding table $\bm{E} = [ \bm{E}_{\mathcal{U}};  \bm{E}_{\mathcal{I}}] \in \mathbb{R}^{d\times |\mathcal{V}|}$, where $\mathcal{V} = \mathcal{U}\cup\mathcal{I}$. Note that the embedding table $\bm{E}$ serves as a starting state for the inference of temporal user/item embeddings. During the training process, $ \bm{E}$ will be 
optimized.

\subsubsection{Continuous-Time Embedding}
The continuous time encoding~\cite{ye2020time,xu2019self} behaves as a function that maps those scalar timestamps into vectors, i.e., $\Phi:T \mapsto \mathbb{R}^{d_{T}}$, where $T \in \mathbb{R}^{+}$. 
Based on previous SR models~\cite{ye2020time,li2020time,wu2020deja}, time span plays a vital component in expressing the temporal effects and uncovering sequential patterns.
The time encoding function embeds timestamps into vectors so as to represent the time span as the dot product of corresponding encoded time embeddings. 
Therefore, we define the temporal effects as a function of time span in continuous time space:
given a pair of interactions $(u, i, t_1)$ and $(u, j, t_2)$ of the same user, the \textbf{temporal effect} is defined as a function $\psi(t_1-t_2)\mapsto \mathbb{R}$, which is expressed as a kernel value of the time embeddings of $t_1$ and $t_2$:
\begin{equation}
\label{eqn:temporal_effects}
    \psi(t_1-t_2) = \mathcal{K}(t_1,t_2) =  \Phi(t_1)\cdot\Phi(t_2),
\end{equation}
where $\mathcal{K}$ is the temporal kernel and $\cdot$ denotes the dot product operation. The temporal effect $\psi(t_1-t_2)$ measures the temporal correlation between two timestamps.
Moreover, 
the time encoding function should be generalized to any unseen timestamp such that any time span not found in training data can still be inferred by the encoded time embeddings.
 Unlike modeling the absolute time difference like~\cite{li2020time}, representing temporal effects as a kernel is generalized to any timestamp as it models the time representations directly. Therefore, the temporal effect of any pair of timestamps can be inductively inferred by the dot product of time representations. 
Eq.~(\ref{eqn:temporal_effects}) can be achieved by a continuous and translation-invariant kernel $\mathcal{K}(t_1, t_2)$ based on Bochner’s Theorem~\cite{loomis2013introduction}. By explicitly representing the temporal features, the temporal embedding is:  
\begin{equation}\label{eq:mapping_function}
    \Phi(t)\mapsto \sqrt{\frac{1}{d_T}}\left[\cos(\omega_1t),\sin(\omega_1t),\dots,\cos(\omega_{d_T}t),\sin(\omega_{d_T}t)\right]^{\top},
\end{equation}
where $\bm{\omega}=\left[\omega_1, \dots, \omega_{d_T}\right]^{\top}$ are learnable and $d_T$ is the dimension.


\subsection{Temporal Collaborative Transformer}
Next, we present the novel \layername layer of \modelname.
We intend to address two strengths of a \layername layer: (1) constructing information from both user/item embeddings and temporal embedding, which explicitly characterizes temporal effects of the correlations;
(2) a collaborative attention module, which advances existing self-attention mechanism by modeling the importance of user-item interactions, which is thus able to explicitly recognize collaborative signals.  
To achieve this, we first present the information construction and aggregation from a user perspective. 
Then, we introduce a novel collaborative attention mechanism to infer importance of interactions. 
Finally, we demonstrate how to generalize to items.

\subsubsection{Information Construction}
We construct input information of each \layername layer as the combination of long term node embeddings and time embeddings. As such, we can unify temporal information and collaborative signals.
In particular, the \textit{query} input information at the $l$-th layer for user $u$ at time $t$ is:
\begin{equation}
\label{eq:query_information}
    \begin{split}
        \bm{h}^{(l-1)}_u(t) &= \bm{e}^{(l-1)}_u(t) \| \Phi(t),
    \end{split}
\end{equation}
where $l=1,2,\dots,L$. $\bm{h}_{u}^{(l-1)}(t)\in\mathbb{R}^{d+d_T}$ is the information for $u$ at $t$, $\bm{e}^{(l-1)}_u(t)\in\mathbb{R}^{d}$ is the temporal embedding of $u$, and $\Phi(t)\in\mathbb{R}^{d_T}$ denotes the time vector of $t$. $\|$ denotes the concatenation operation. Other operations including summation are possible. However, we use concatenation for simplicity. It also provides intuitive interpretation in the attention, as shown in Eq.~(\ref{eq:att_interpret}).  Note that when $l=1$, it is the first \layername layer. The temporal embedding $\bm{e}_u^{(0)}(t)=\bm{E}_u$, i.e., the long-term user embedding. When $l>1$, the temporal embedding is generated from the previous \layername layer.

In addition to the query node itself, to 
we also propagate temporal collaborative information from its neighbors.
We randomly sample $S$ different interactions of $u$ before time $t$ as $\mathcal{N}_u(t)=\{(i, t_s)|(u, i, t_s)\in\mathcal{E}_{t}\text{ and }t_s <t\}$. The input information at the $l$-th layer for each $(i,t_s)$ pair is:
\begin{equation}
\label{eq:neighbor_information}
    \begin{split}
        \bm{h}^{(l-1)}_i(t_s) &= \bm{e}^{(l-1)}_i(t_s) \| \Phi(t_s),
    \end{split}
\end{equation}
where $\bm{h}_i(t_s)$ is the information for item $i$ at $t_s$, $\bm{e}_i(t_s)$ denotes the temporal embedding of $i$ at $t_s$. Again, note that when $l=1$, $\bm{e}_i^{(0)}(t_s)=\bm{E}_i$, i.e., the long-term item embedding. When $l>1$, the temporal embedding is output from the previous \layername layer.

\subsubsection{Information Propagation}
After constructing the information, we propagate the information of sampled neighbors $\mathcal{N}_u(t)$ to infer the temporal embeddings. 
Since the neighbors are involving with time $t$, in this way, we can unify the sequential patterns with temporal collaborative signals.
We compute the linear combination of the information from all sampled interactions as:
\begin{equation}
\label{eq:propagation}
    \bm{e}^{(l)}_{\mathcal{N}_u}(t) = \sum_{(i, t_s)\in N_u(t)}
    \pi_t^{u}(i, t_s)\bm{W}^{(l)}_{v}\bm{h}^{(l-1)}_i(t_s),
\end{equation}
where $\pi_t^{u}(i, t_s)$\footnote{$\pi_t^{u}(i, t_s)$ also has a superscript of the layer number $l$, which is ignored for simplicity.} denotes the importance of an interaction $(u,i,t_s)$ and $\bm{W}_{v}\in\mathbb{R}^{d \times (d+d_T)}$ is the linear transformation matrix.  $\pi_t^{u}(i, t_s)$ represents the impact of a historical interaction $(u,i,t_s)$ to the temporal inference of $u$ at time $t$, which is calculated by the temporal collaborative attention.

\subsubsection{Temporal Collaborative Attention}
\label{subsubsec:self_att}
We adopt the novel temporal collaborative attention mechanism to measure the weights $\pi_t^{u}(i, t_s)$, which considers both neighboring interactions and the temporal information on edges. Both factors contribute to the importance of historical interactions. Thus, it is a better mechanism to capture temporal collaborative signals than self-attention mechanism that
only models item-item correlations. The attention weight $\pi_t^{u}(i, t_s)$ is formulated as follows:
\begin{equation}\label{eq:dot_attention}    
    \pi_t^{u}(i, t_s) = \frac{1}{\sqrt{d+d_T}} \left(\bm{W}^{(l)}_{ k}\bm{h}^{(l-1)}_{i}(t_s)\right)^{\top} \bm{W}^{(l)}_{q}\bm{h}^{(l-1)}_u(t),
\end{equation}
where $\bm{W}^{(l)}_{k}$ and $\bm{W}^{(l)}_{q}$ are both linear transformation matrices, and the factor $\frac{1}{\sqrt{d+d_T}}$ protects the dot-product from growing large with high dimensions. We adopt dot-product attention because if we ignore transformation matrices and the scalar factor, based on Eq.~(\ref{eq:query_information}) and Eq.~(\ref{eq:neighbor_information}), the right-hand side of Eq.~(\ref{eq:dot_attention}) can be rewritten as:
\begin{equation}
\label{eq:att_interpret}
\begin{split}
     \bm{e}^{(l-1)}_u(t) \cdot \bm{e}^{(l-1)}_i(t_s) + \Phi(t)\cdot\Phi(t_s),
\end{split}
\end{equation}
where the first term denotes the user-item collaborative signal, and the second term models the temporal effect according to Eq.~(\ref{eqn:temporal_effects}). With more stacked layers, collaborative signals and temporal effects are entangled and tightly connected. Hence, the dot-product attention can characterize impacts of temporal collaborative signals.

Hereafter, we normalize the attention weights across all sampled interactions by employing a softmax function:
\begin{equation}\label{eq:softmax}
    \pi_t^{u}(i, t_s) = \frac{\text{exp}\left(\pi_t^{u}(i, t_s)\right)}{\sum_{(i', t_s')\in N_u(t)}\text{exp}\left(\pi_t^{u}(i', t_s')\right)}.
\end{equation}
Moreover, the computation is implemented by packing the information of all sampled interactions. To be more specific, we stack the information (Eq.~(\ref{eq:neighbor_information})) of all sampled interactions as a matrix $\bm{H}^{(l-1)}_{\mathcal{N}_{u}}(t)\in\mathbb{R}^{(d+d_T)\times S}$, as illustrated\footnote{In figure~\ref{fig:architect_agg}, an embedding is a row vector, while in notations, it is a column vector.} in Figure~\ref{fig:architect_agg}. We denote $\bm{K}^{(l-1)}_{u}(t)=\bm{W}^{(l)}_{k}\bm{H}^{(l-1)}_{\mathcal{N}_{u}}(t)$, $\bm{V}^{(l-1)}_{u}(t)=\bm{W}^{(l)}_{v}\bm{H}^{(l-1)}_{\mathcal{N}_{u}}(t)$ and $\bm{q}^{(l-1)}_{u}(t)=\bm{W}^{(l)}_{q}\bm{h}^{(l-1)}_{u}(t)$, which are respectively the key, value and query input for the temporal collaborative attention module. We illustrate this in Figure~\ref{fig:architect_agg} as green blocks. For simplicity and without ambiguity, we ignore the superscripts and time $t$ and combine Eq.~(\ref{eq:dot_attention}) and Eq.~(\ref{eq:softmax}). Then, we can rewrite the Eq.~(\ref{eq:propagation}) as:
\begin{equation}\label{eq:dot-product attention}
    \bm{e}_{\mathcal{N}_u} = \bm{V}_{u}\cdot \text{Softmax}\left(\frac{\bm{K}^{\top}_u\bm{q}_u}{\sqrt{d+d_T}}\right) ,
\end{equation}
which is in the form of dot-product attention in Transformer~\cite{vaswani2017attention}. Therefore, we can safely apply the multi-head attention operation and concatenate the output from each head as the information for aggregation, which is presented in Figure~\ref{fig:architect_agg}. 
Note that our attention is not a self-attention but a temporal collaborative attention, which jointly models user-item interactions and temporal information. 

\subsubsection{Information Aggregation}
To output the temporal node embedding, the final step of a \layername layer is to aggregate the query information in Eq.~(\ref{eq:query_information}) and the neighbor information in Eq.~(\ref{eq:propagation}). We concatenate and send them to a feed-forward neural network~(FFN):
\begin{equation}
    \begin{split}
         \bm{e}^{(l)}_u(t) &=   \text{FFN}\left( \bm{e}^{(l)}_{\mathcal{N}_u}(t) \|  \bm{h}^{(l-1)}_u(t)\right),
    \end{split}
\end{equation}
where $\bm{e}^{(l)}_u(t)$ is the temporal embedding of $u$ at $t$ on $l$-th layer, and FFN consists of two linear transformation layers with a ReLU activation function in between~\cite{vaswani2017attention}. The output temporal embedding $\bm{e}^{(l)}_u(t)$ can either be sent to the next layer or output as the final temporal node embedding for prediction.

\subsubsection{Generalization to items} Though we only present the \layername layer from the user query perspective, it is analogous if the query is an item at a specific time. We only need to alternate the user query information to the item query information, and change the neighbor information in Eq.~(\ref{eq:neighbor_information}) and Eq.~(\ref{eq:propagation}) accordingly as user-time pairs. Then, we can make an inference of the temporal embedding of item $i$ at time $t$ as $\bm{e}^{(l)}_i(t)$, which is sent to the next layer.

\subsection{Model Prediction}
The \modelname model consists of $L$ \layername layers. For each test triplet $(u, i, t)$, it yields temporal embeddings for both $u$ and $i$ at $t$ on the last \layername layer, denoting as $\bm{e}^{(L)}_u(t)$ and $\bm{e}^{(L)}_i(t)$, respectively.  Then, the prediction score is:
\begin{equation}
    \label{eq:pred_uit}
    r(u,i,t) =  
    \bm{e}^{(L)}_u(t)\cdot \bm{e}^{(L)}_i(t)
    ,
\end{equation}
where $r(u,i,t)$ denotes the score to recommend $i$ for $u$ at time $t$. With the generalized continuous-time embeddings and the proposed \layername layers, we can generalize and infer user\slash item embeddings at any timestamp, thus making multiple steps recommendation feasible while existing work only predicts next item. Recall that based on the Definition~\ref{def:ct_seq_rec}, we recommend each user a ranking list of items at the given timestamp. Therefore, we use Eq.~(\ref{eq:pred_uit}) to calculate scores of all candidate items and sort them by scores.

\subsection{Model Optimization}
To learn the model parameters, we use the pairwise BPR loss~\cite{rendle2009bpr}, which is widely used for top-N recommendation. The pairwise BPR loss assumes that the observed implicit feedback items have greater prediction scores than those unobserved and is also designed for ranking based top-N recommendation. The objective function is:
\begin{equation}
    \mathcal{L}_{bpr} = \sum_{(u,i,j,t)\in \mathcal{O}_T} -\text{log}\sigma\left(r(u,i,t) - r(u,j,t)\right) + \lambda||\Theta||_2^2,
\end{equation}
where $\mathcal{O}_T$ denotes the training samples, $\Theta$ includes all learnable parameter, and $\sigma(\cdot)$ is a sigmoid function. The training samples $\mathcal{O}_T=\left\{(u,i,j,t)|(u,i,t) \in \mathcal{E}_T, j\in\mathcal{I} \setminus \mathcal{I}_u(t)\right\}$, where the positive interaction $(u,i,t)$ comes from the edge set $\mathcal{E}_T$ of CTBG, the negative item $j$ is sampled from unobserved items $\mathcal{I} \setminus \mathcal{I}_u(t)$ of user $u$ at timestamp $t$; $\Theta$ includes long-term embedding $E$, time embedding parameter $\omega$, and all linear transformation matrices. The loss is optimized via mini-batch Adam~\cite{DBLP:journals/corr/KingmaB14} with adaptive learning rate. Alternatively, we can optimize the model with a Binary Cross Entropy~(BCE) loss as:
\begin{equation}
    \mathcal{L}_{bce} = \sum_{(u,i,j,t)\in \mathcal{O}_T} \log\sigma\left(r(u,i,t)\right) + \log\sigma\left(1 - r(u,j,t)\right) + \lambda||\Theta||_2^2,
\end{equation}
which is compared with BPR loss in experiments.

\section{Experiments}
In this section, we present the experimental setups and results to demonstrate the effectiveness of \modelname.
The experiments answer the following Research Questions~(RQs):
\begin{itemize}
    \item \textbf{RQ1}: Does \modelname yield better recommendation?
    \item \textbf{RQ2}: How do different hyper-parameters (\textit{e.g.,} number of neighbors $S$, etc.) affect the performance of \modelname?
    \item \textbf{RQ3}: How do different modules (\textit{e.g.,} temporal collaborative attention, etc.) affect the performance of \modelname?
    \item \textbf{RQ4}: Can \modelname effectively unify sequential patterns and temporal collaborative signals? (Reveal temporal correlations)
   
\end{itemize}

\subsection{Datasets}
We conduct our experiments on four Amazon review datasets~\cite{mcauley2015image} and MovieLens ML-100K dataset~\cite{harper2015movielens}. The Amazon datasets are collected from different domains\footnote{https://jmcauley.ucsd.edu/data/amazon/}, from the Amazon website during May 1996 to July 2014. The Movie Lens dataset is collected from September 19th, 1997 through April 22nd, 1998. We use Unix timestamps on all datasets.
For each dataset, we chronologically split for train\slash validation\slash test in 80\%/10\%/10\% ratio based on the interaction timestamps. 
More details, such as data descriptions and statistics, are presented in the Table~\ref{tab:stat_data}. We can find amazon datasets are much sparser and their time spans are much longer compared with ML-100K dataset. 
For Amazon related datasets, the time intervals of successive interactions are typically in days, while ML-100k has shorter time intervals, ranging from seconds to days.

\begin{table}[h]
\begin{threeparttable}
\caption{Statistics of datasets.}
\label{tab:stat_data}
\begin{tabular}{lccccc}
\toprule
\multicolumn{1}{c}{Dataset} &
  \multicolumn{1}{c}{Toys} &
  \multicolumn{1}{c}{Baby} &
  \multicolumn{1}{c}{Tools} &
  \multicolumn{1}{c}{Music} &
  \multicolumn{1}{c}{ML100K} \\ \midrule
{\#Users} & 17,946  & 17,739  & 15,920   & 4,652     & 943      \\
{\#Items} & 11,639  & 6,876   & 10,043   & 3,051    & 1,682    \\
{\#Edges} & 154,793 & 146,775 & 127,784  & 54,932  & 48,569   \\
{\#Train} & 134,632 & 128,833 & 107,684  & 51,765  & 80,003   \\
{\#Valid} & 11,283  & 10,191  & 10,847   & 2,183   & 1,516    \\
{\#Test}  & 8,878   & 7,751   & 9,253    & 984     & 1,344    \\
Density   & 0.07\%  & 0.12\%  & 0.08\%   & 0.38\%   & 6.30\%   \\
Avg. Int. & 85 days & 61 days & 123 days & 104 days & 4.8 hours \\
\bottomrule

\end{tabular}%
\begin{tablenotes}
      \small
      \item  ``Av. Int.'' denotes average time interval.
    \end{tablenotes}
\end{threeparttable}
\end{table}
\vspace{-5pt}

\subsection{Experimental Settings}
\begin{table*}[]
\caption{Overall Performance w.r.t. Recall@\{10,20\} and MRR.}
\label{tab:overall_perf}
\resizebox{1\textwidth}{!}{%
\begin{tabular}{lcccccccccccccc}
\toprule
Datasets &
  Metric &
  \multicolumn{1}{c}{BPR} &
  \multicolumn{1}{c}{\text{LightGCN}} &
  \multicolumn{1}{c}{SR-GNN} &
  \multicolumn{1}{c}{GRU4Rec} &
  \multicolumn{1}{c}{Caser} &
  \multicolumn{1}{l}{SSE-PT} &
  BERT4Rec &
  \multicolumn{1}{c}{SASRec} &
  \multicolumn{1}{l}{TiSASRec} &
  \multicolumn{1}{c}{CDTNE} &
  \multicolumn{1}{c}{{\color[HTML]{333333} \modelname}} &
  \multicolumn{1}{c}{Improv.} \\ 
  \toprule
 &
  Recall@10 &
  0.0021 &
  0.0016 &
  0.0020 &
  0.0274 &
  0.0138 &
  0.1213 &
  0.1273 &
  {\ul 0.1452} &
  0.1361 &
  0.0016 &
  \textbf{0.3650} &
  0.2198 \\
 &
  Recall@20 &
  0.0036 &
  0.0026 &
  0.0033 &
  0.0288 &
  0.0238 &
  0.1719 &
  0.1865 &
  {\ul 0.2044} &
  0.1931 &
  0.0045 &
  \textbf{0.3714} &
  0.1670 \\
\multirow{-3}{*}{Toys} &
  MRR &
  0.0024 &
  0.0018 &
  0.0018 &
  0.0277 &
  0.0082 &
  0.0595 &
  0.0643 &
  {\ul 0.0732} &
  0.0671 &
  0.0025 &
  \textbf{0.3661} &
  0.2929 \\ \midrule
 &
  Recall@10 &
  0.0028 &
  0.0036 &
  0.0030 &
  0.0036 &
  0.0077 &
  0.0911 &
  0.0884 &
  0.0975 &
  {\ul 0.1040} &
  0.0218 &
  \textbf{0.2235} &
  0.1195 \\
 &
  Recall@20 &
  0.0039 &
  0.0045 &
  0.0062 &
  0.0048 &
  0.0193 &
  0.1418 &
  0.1634 &
  0.1610 &
  {\ul 0.1662} &
  0.0292 &
  \textbf{0.2295} &
  0.0663 \\
\multirow{-3}{*}{Baby} &
  MRR &
  0.0019 &
  0.0024 &
  0.0024 &
  0.0028 &
  0.0071 &
  0.0434 &
  0.0511 &
  0.0455 &
  {\ul 0.0521} &
  0.0157 &
  \textbf{0.2147} &
  0.1626 \\ \midrule
 &
  Recall@10 &
  0.0023 &
  0.0021 &
  0.0051 &
  0.0048 &
  0.0077 &
  0.0775 &
  {\ul 0.1296} &
  0.0913 &
  0.0946 &
  0.0186 &
  \textbf{0.2457} &
  0.1161 \\
 &
  Recall@20 &
  0.0036 &
  0.0035 &
  0.0092 &
  0.0059 &
  0.0161 &
  0.1155 &
  {\ul 0.1784} &
  0.1337 &
  0.1356 &
  0.0271 &
  \textbf{0.2559} &
  0.0775 \\
\multirow{-3}{*}{Tools} &
  MRR &
  0.0026 &
  0.0023 &
  0.0028 &
  0.0051 &
  0.0068 &
  0.0419 &
  {\ul 0.0628} &
  0.0460 &
  0.0480 &
  0.0203 &
  \textbf{0.2468} &
  0.1840 \\ 
  \midrule
 &
  Recall@10 &
  0.0122 &
  0.0142 &
  0.0051 &
  0.0549 &
  0.0183 &
  0.0915 &
  0.1352 &
  {\ul 0.1372} &
  {\ul 0.1372} &
  0.0071 &
  \textbf{0.5935} &
  0.4563 \\
 &
  Recall@20 &
  0.0152 &
  0.0183 &
  0.0092 &
  0.0589 &
  0.0346 &
  0.1494 &
  0.2093 &
  {\ul 0.2094} &
  0.1951 &
  0.0163 &
  \textbf{0.5986} &
  0.3892 \\
\multirow{-3}{*}{Music} &
  MRR &
  0.0057 &
  0.0064 &
  0.0028 &
  0.0540 &
  0.0106 &
  0.0423 &
  {\ul 0.0824} &
  0.0768 &
  0.0681 &
  0.0037 &
  \textbf{0.3820} &
  0.2996 \\ 
  \midrule
   \multicolumn{1}{l}{} &
  Recall@10 &
  0.0461 &
  0.0565 &
  0.0045 &
  0.0996 &
  0.0246 &
  0.1079 &
  0.1116 &
  0.09450 &
  {\ul 0.1332} &
  0.0350 &
  \textbf{0.3118} &
  0.1786 \\
\multicolumn{1}{l}{} &
  Recall@20 &
  0.0766 &
  0.0960 &
  0.0060 &
  0.1168 &
  0.0417 &
  0.1801 &
  0.1786 &
  0.1808 &
  {\ul 0.2232} &
  0.0536 &
  \textbf{0.3252} &
  0.1020 \\
\multicolumn{1}{l}{\multirow{-3}{*}{ML100k}} &
  MRR &
  0.0213 &
  0.0252 &
  0.0012 &
  {\ul 0.0938} &
  0.0147 &
  0.0519 &
  0.0600 &
  0.0448 &
  0.0605 &
  0.0162 &
  \textbf{0.2416} &
  0.1478 \\
  \bottomrule
\end{tabular}%
}
\end{table*}

\begin{figure*}[ht]
     \centering
     \begin{subfigure}[b]{0.19\textwidth}
         \centering
         \includegraphics[width=1\textwidth]{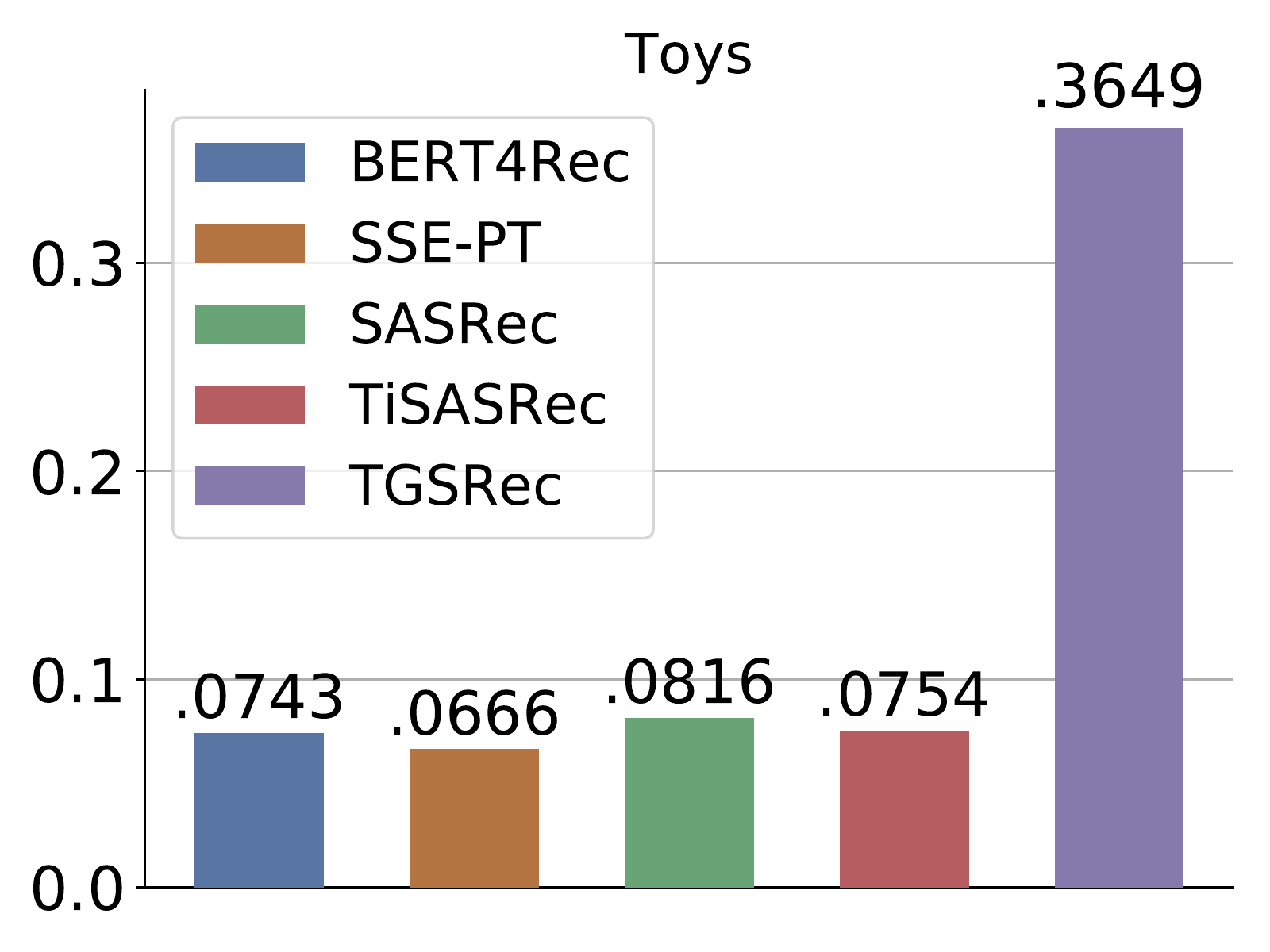}
         \caption{NDCG@10 in Toys}
         \label{fig:ndcg_toy}
     \end{subfigure}\hfill
     \begin{subfigure}[b]{0.19\textwidth}
         \centering
         \includegraphics[width=1\textwidth]{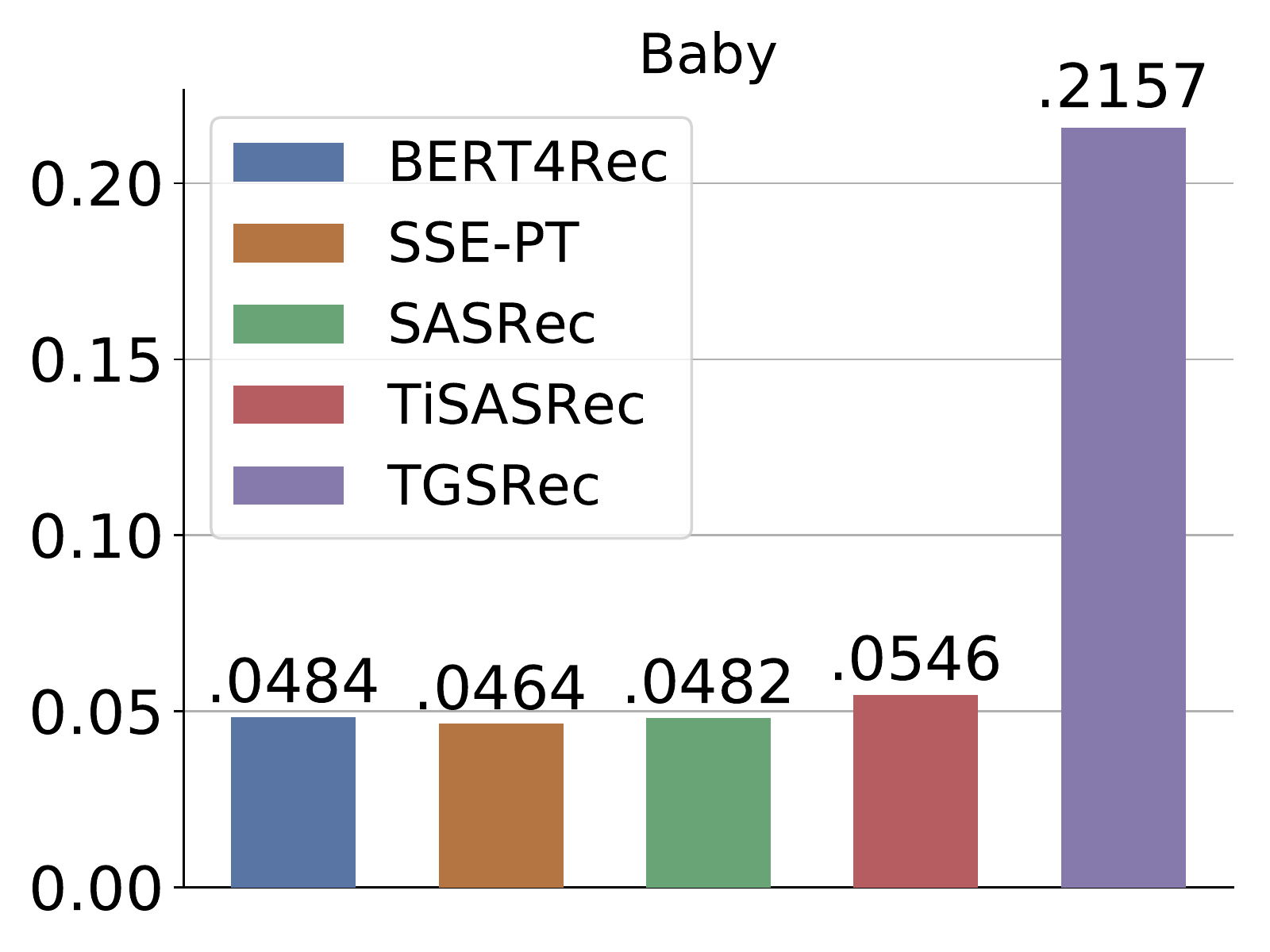}
         \caption{NDCG@10 in Baby}
         \label{fig:ndcg_baby}
     \end{subfigure}\hfill
     \begin{subfigure}[b]{0.19\textwidth}
         \centering
         \includegraphics[width=1\textwidth]{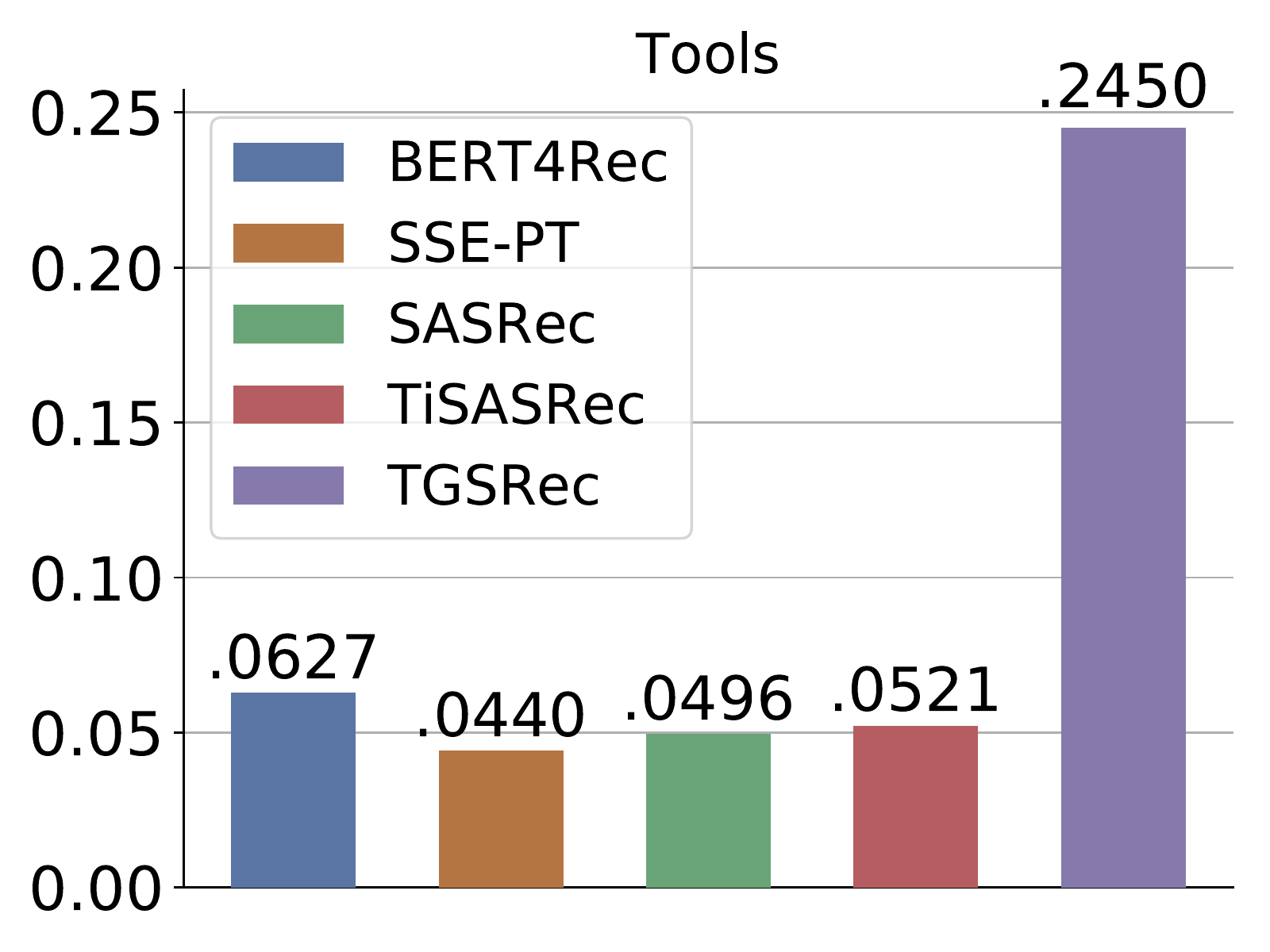}
         \caption{NDCG@10 in Tools}
         \label{fig:ndcg_tools}
     \end{subfigure}\hfill
     \begin{subfigure}[b]{0.19\textwidth}
         \centering
         \includegraphics[width=1\textwidth]{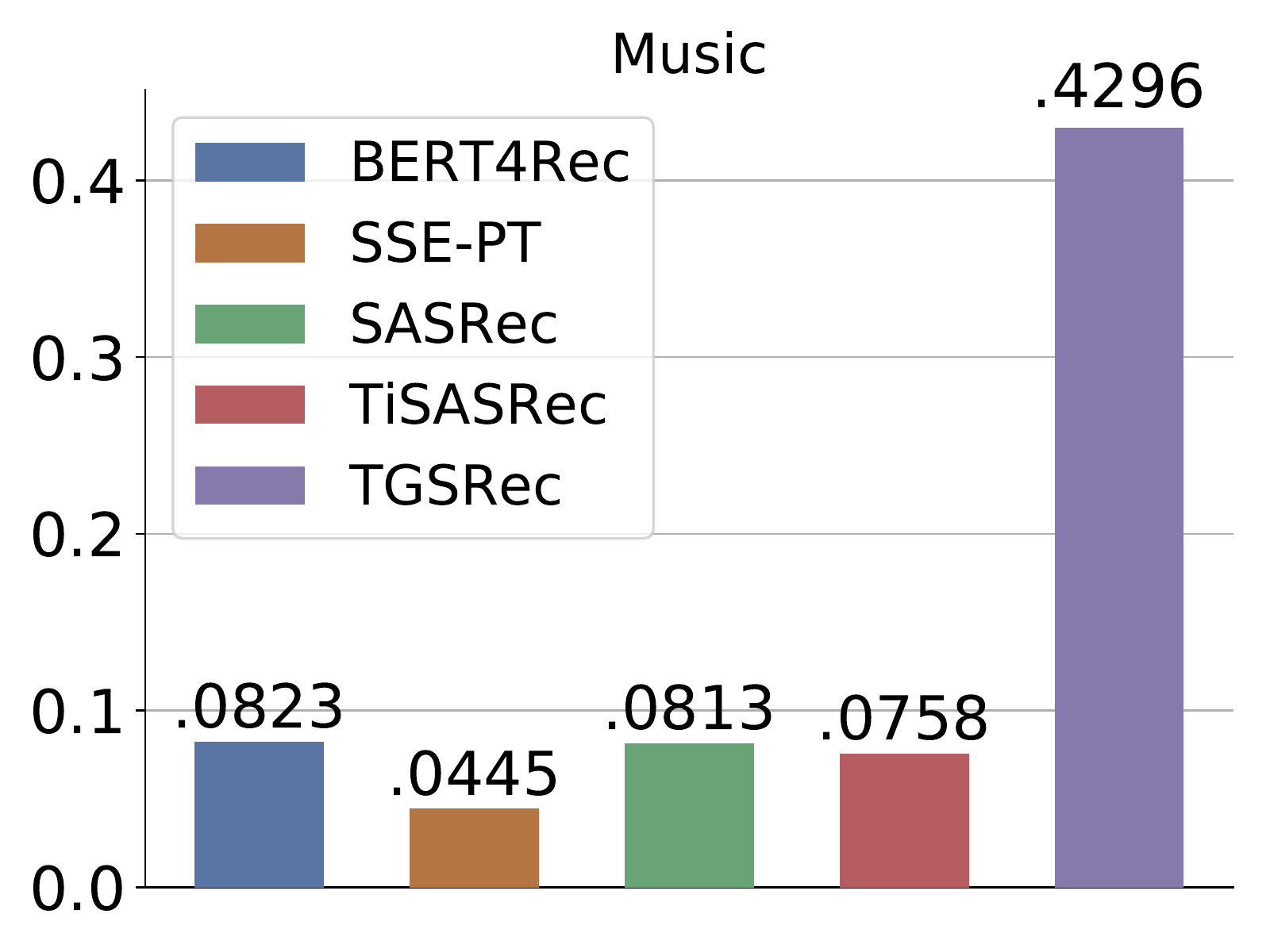}
         \caption{NDCG@10 in Music}
         \label{fig:ndcg_music}
     \end{subfigure}\hfill
     \begin{subfigure}[b]{0.19\textwidth}
         \centering
         \includegraphics[width=\textwidth]{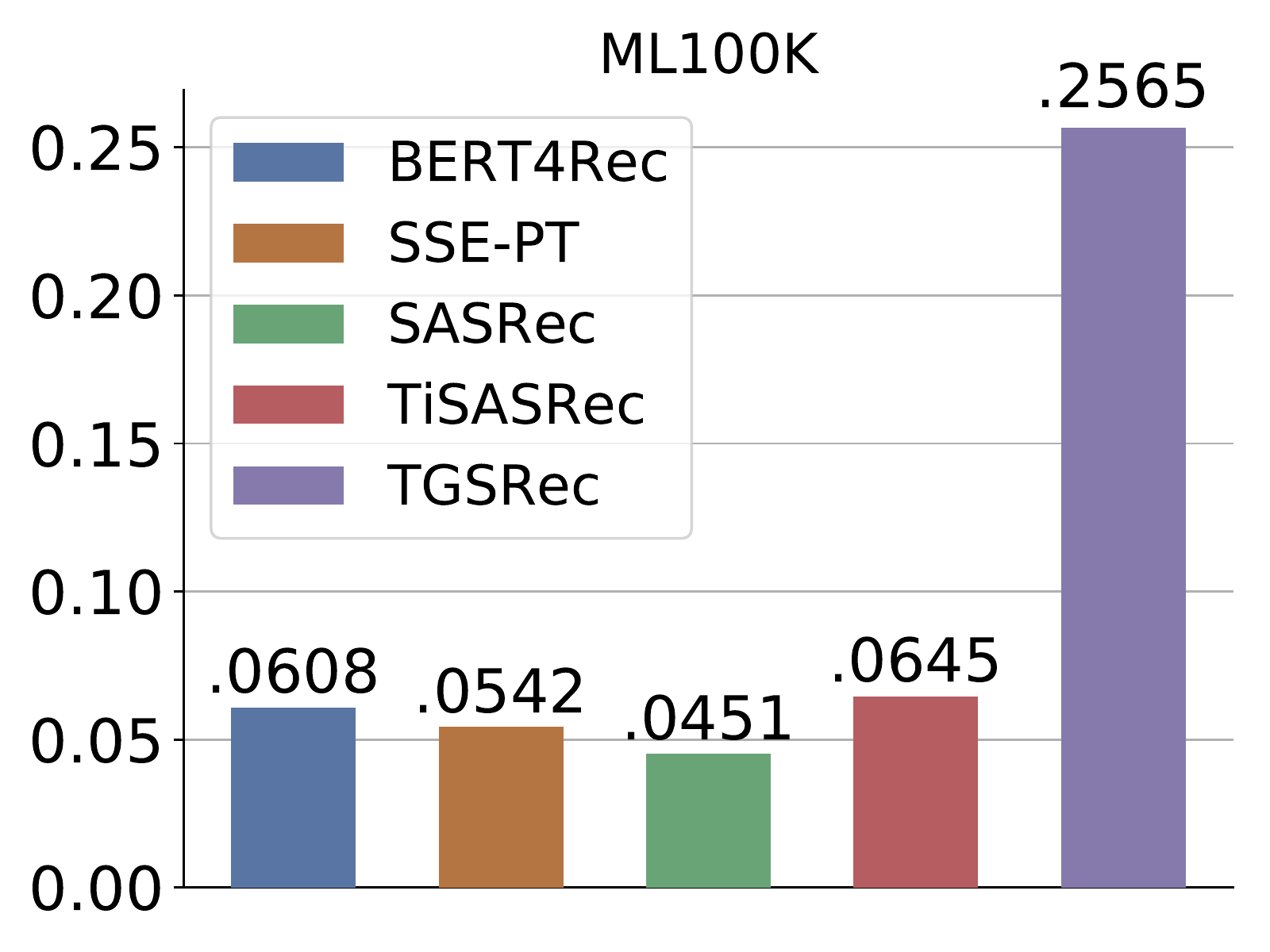}
         \caption{NDCG@10 in ML100K}
         \label{fig:ndcg_office}
     \end{subfigure}
     \vspace{-3mm}
        \caption{NDCG@10 Performance in all Datasets. We ignore other methods because of their low values.}
        \label{fig:ndcg_alldata}
\end{figure*}

\subsubsection{Baselines}
We compared \modelname with the state-of-the-art methods in three different groups.
\textbf{Static models}: Static models ignore the temporal information and generate static user\slash item embeddings for a recommendation. We compare with the most standard baseline BPRMF~\cite{rendle2009bpr}, and also compare with a recent GNN-based model LightGCN~\cite{he2020lightgcn}.
\textbf{Temporal models}: We compare some relevant temporal methods, such as CTDNE~\cite{nguyen2018continuous} and one recent model TiSASRec~\cite{li2020time}, which utilize time information. 
We also try to compare with JODIE~\cite{kumar2019predicting}. However, we do not report it because has out-of-memory errors on most datasets.
\textbf{Transformer-based SR models}: Since our model is built upon the transformer, we mainly focus on comparing with the recent transformer-based SR methods, which are SASRec~\cite{kang2018self}, BERT4Rec \cite{sun2019bert4rec}, SSE-PT~\cite{ssept20wu}, and TiSASRec~\cite{li2020time}. 
\textbf{Other SR models}: In addition, we also compare with other type of SR models, i.e., FPMC~\cite{rendle2010factorizing}, GRU4Rec~\cite{hidasi2015session}, Caser~\cite{tang2018personalized}, and SR-GNN~\cite{wu2019session}, for comprehensive study.

For each testing interaction $(u, i, t_{test})$, our continuous-time sequential recommendation setting allows models to use any history interactions $\{(u, i, t)|t<t_{test}\}$ during the prediction stage, regardless of whether the historical interactions are in training portion, validation part or even in testing set. However, all parameters of models are only learned from the training data.

We implement \modelname with Pytorch in a Nvidia 1080Ti GPU. We grid search all parameters and report the test performance based on the best validation results. For all models, we search for the dimensions of embeddings $d$ in range of $[8, 16, 32, 64]$ and we tune the learning rate in $[10^{-2},10^{-3},10^{-4}]$, search the L2 regularization weight from $[5\times 10^{-1}, 10^{-1}, 10^{-2}, 10^{-3}]$. For sequential methods, we search the maximum length of sequence in $[50, 100]$, number of layers from $[1,2,3]$, and number of heads in $[1,2,4]$.

\subsubsection{Evaluation Protocol} 
All models will generate a ranking list of items for each testing interaction. Each evaluation metric is averaged over the total number of interactions as the final reported result. 
In order to accelerate the evaluation, we sample 1,000 negative items for evaluation instead of full set of negative items. For each interaction $(u,i,t)$ in validation\slash test sets, we treat items that $u$ has no interactions with before $t$ as negative items. Regarding the sampling bias for evaluation~\cite{krichene2020sampled}, we apply the unbiased estimator in~\cite{krichene2020sampled} to correct the sampled ranks. We evaluate the \textit{top-N} recommendation performance by standard ranking-based evaluation metrics \textit{Recall@N}, \textit{NDCG@N}, and \textit{Mean Reciprocal Rank~(MRR)}. 
We set N to be either 10 or 20 for a comprehensive comparison.

\vspace{-8pt}
\subsection{Performance Comparison (RQ1)}
We compare the performance of all models and demonstrate the superiority of \modelname. We report the Recall and MRR of all models in Table~\ref{tab:overall_perf}. Additionally, we visualize the comparisons of NDCG in Figure~\ref{fig:ndcg_alldata}. We have the following observations:
\begin{itemize}
    \item \modelname consistently and significantly outperforms all baselines in all datasets. In particular, for absolute performance improvement gains relative to the 2nd best, \modelname achieves $22.51\%$, $16.90\%$ and $22.15\%$ absolute gains at recall@10, recall@20, and MRR, respectively. \modelname also significantly outperforms others in NDCG, as shown in Figure~\ref{fig:ndcg_alldata}. Several factors together determine the superiority of \modelname: (1)~\modelname captures temporal collaborative signals; (2)~\modelname explicitly expresses temporal effects; and (3)~\modelname stacks multiple \layername layers to propagate the information. 
    \item Those static methods achieve the worst performance among all models. A simple GRU4Rec even performs 10 times better than them. This indicates that static embeddings fail to utilize the temporal information, limiting its recommendation ability in SR. Thus, it is important to model dynamics.
    \item The CDTNE performs better than Caser and GRU4Rec in Tools and Baby datasets. This suggests the benefit of modeling temporal information with a graph. But it is still much worse than those transformer-based methods, which again proves the strength of transformer in encoding sequences. We also notice the poor performance of SR-GNN. We analyze the data and find time intervals between successive interactions vary a lot. Since SR-GNN is originally designed for session-based sequences, it is not suitable for SR with a long time span.
    \item The transformer-based SR methods consistently outperform all other types of baselines, which demonstrates the effectiveness of using transformer structure to encode sequence. Among them, TiSASRec is better than SASRec on two datasets, which proves the effectiveness of using time information. 
    But it is still far worse than \modelname. The reason is twofold. One is that only the interval information is not enough to unify the temporal information with sequential patterns. The other is that the proposed temporal collaborative attention in \layername layer captures more precise and generalized temporal effects.
    We find that BERT4Rec is better than the other baselines on the Tools dataset but not better on other datasets. Since the main difference between BERT4Rec and SASRec is the bi-directional sequence encoding, it may break causal relations among items within a sequence. 
    The \modelname performs much better than SR models, showing the necessity of unifying sequential patterns and temporal collaborative signals.

\end{itemize}

\begin{table}[]
\caption{ Ablation analysis (Recall@10) on five datasets. Bold score indicates performance better than the default version, while $\downarrow$ indicates a performance drop more than 50\%.}
\begin{tabular}{@{}lrrrrr@{}}
\toprule
Architecture                  & Toys   & Baby   & Tools           & Music  & ML100K \\ \midrule
(0) Default                   & \textbf{0.3649} & \textbf{0.2235} & \textbf{0.3623}          & \textbf{0.5935} & 0.3118\\
\midrule
(1) Mean                & 0.0027$\downarrow$ & 0.0210$\downarrow$ & 0.0055$\downarrow$          & 0.0051$\downarrow$ & 0.0647$\downarrow$\\
(2) LSTM                & 0.0991$\downarrow$ & 0.1237 & 0.1266$\downarrow$          & 0.3740  & 0.3088 \\
\midrule
(3) Fixed $\omega$ & 0.0854$\downarrow$       &   0.0944$\downarrow$     &    0.0910$\downarrow$             &   0.3679  & 0.2789 \\
(4) Position & 0.0380$\downarrow$       & 0.0243$\downarrow$   &  0.0209$\downarrow$      &  0.0742$\downarrow$   & 0.0878$\downarrow$     \\
(5) Empty & 0.0139$\downarrow$       & 0.0240$\downarrow$  &   0.0018$\downarrow$    & 0.0346$\downarrow$ & 0.0603$\downarrow$\\
\midrule

(6) BCELoss                   &  0.2200      &  0.1916      &    0.1763$\downarrow$             & 0.4624 & \textbf{0.3542}\\
\bottomrule
\end{tabular}%
\label{tab:ablation_study}
\end{table}

\subsection{Parameter Sensitivity (RQ2)}
\begin{figure}
    \centering
    \includegraphics[width=0.45\textwidth]{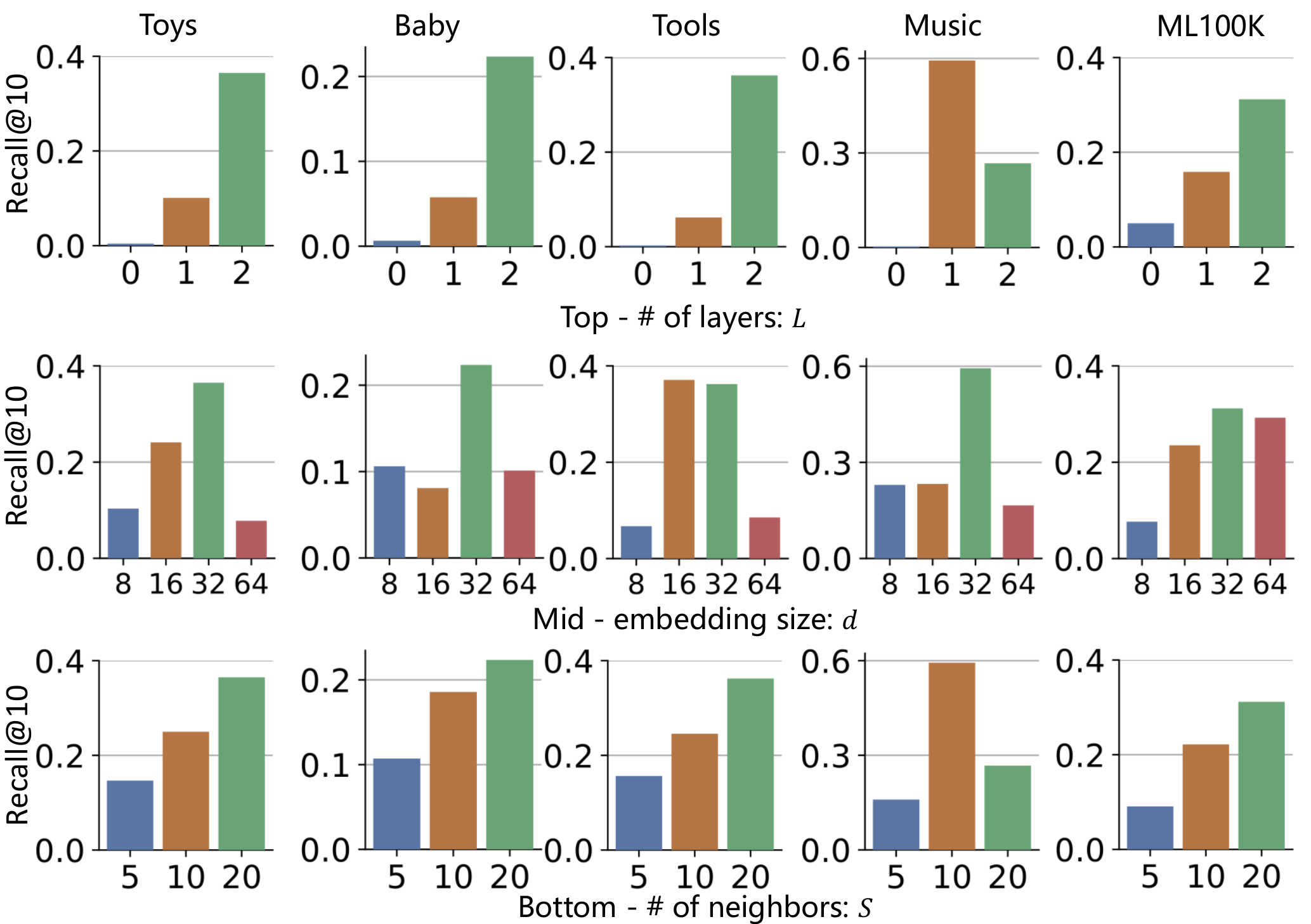}
    \vspace{-3mm}
    \caption{Recall@10 w.r.t. $L,d$ and $S$ on $5$ datasets.}
    \label{fig:hyper}
\end{figure}
In this section, we conduct sensitivity analyses of the hyperparameters of \modelname, including the number of layers $L$, embedding size $d$, and the number of neighbors $S$. The results are reported in Figure~\ref{fig:hyper}.\\
    \noindent \textbf{The number of layers.} The number of \layername layers $L$ is searched from $\{0,1,2\}$. The results are shown in the top row of Figure~\ref{fig:hyper}. When $L=0$, \modelname has no \layername layer, thus unable to infer temporal embeddings. We can observe it performs the worst on all dataset, which justify the benefit of temporal inference. When $L=1$, it makes temporal inference, but without propagation to the next layer. Therefore, it still performs worse than $L=2$ on most datasets. When $L=2$, it can not only make temporal inference, but also propagate the information to capture high-order signals, which alleviates the data sparsity problem. \\
    \noindent \textbf{Embedding size.} The embedding size $d$ of \layername layers is searched from $\{8,16,32,64\}$, which is presented at the mid-row in Figure~\ref{fig:hyper}. We can find that the performance increases as the embedding size enlarges. However, when the embedding size is too large, \textit{e.g.,} $d=64$, the performance drops, which results from the over-fitting problem because of too many parameters. \\
    \noindent \textbf{Number of neighbors.} The number of neighbors $S$ is searched in $\{5,10,20\}$, which is illustrated in the bottom row of Figure~\ref{fig:hyper}. We can observe that \modelname has performance gains on most datasets as the number of neighbors grows. It is because more neighbors can provide more information for encoding both sequences and temporal collaborative signals. 

\subsection{Ablation Study (RQ3 \& RQ4)}\label{sec:ablation}
In this section, we conduct experiments to analyze different components in \modelname. We develop several variants to better understand their effectiveness. Table~\ref{tab:ablation_study} shows the performance w.r.t. Recall@10 of the default \modelname and other variants. We label each row with an index number for quick reference. The default is \modelname with all components and labeled as $0$. We develop the variants by substituting some components, which are temporal collaborative attention (1-2), continuous-time embeddings (3-5), and loss function (6):\\
    \textbf{Temporal collaborative attention.} We replace the proposed temporal collaborative attention of sampled neighbors with a mean pooling or LSTM module, both of which are widely used to encode sequences. Results are labeled as $(1)$ and $(2)$ in Table~\ref{tab:ablation_study}. We can observe that substituting collaborative attention with a mean pooling layer severely spoils the performance. Compared with that, the adoption of LSTM is much better, indicating the necessity of encoding sequential patterns by considering item transitions. However, both of them are worse than the default one, which implies the advantage of temporal collaborative attention in encoding sequences. \\
    \noindent \textbf{Continuous-time embedding.} 
    We use three variants to verify the efficacy of the time mapping function $\Phi$. The first variant is that we sample $\omega$ in Eq.~(\ref{eq:mapping_function}) directly from a normal distribution. The second and third variants replace the $\Phi$ with a learnable positional embedding as in~\cite{kang2018self} and emptying all zeros, respectively.
    The results are labeled as $(3)-(5)$ in Table~\ref{tab:ablation_study}. Because of the better performance of position embedding compared with empty embedding, we can conclude that \modelname has the ability to encode sequential patterns. In addition, we also find that even a fixed $\omega$ to learn the time embedding can significantly outperform the position embedding, indicating the necessity of using the temporal kernel to capture temporal effects in sequences. Moreover, the default version, using a trainable $\omega$, achieves the best performance, which indicates its capacity to learn temporal effects from data.\\
    \noindent \textbf{Loss function.} We also compare BPR loss and BCE Loss, which is labeled as $(6)$ in Table~\ref{tab:ablation_study}. The results indicate that the BCE loss performs inferior to BPR loss, except for the ML100K dataset. This is because BPR loss is optimized for ranking while BCE loss is designed for binary classification.\\

\vspace{-8pt}
\subsection{Temporal Correlations (RQ4)}
Though we have already indicated the answer of RQ4 in Sec.~\ref{sec:ablation}, this section also conducts detailed analyses of the temporal  correlation within sequences to directly answer \textbf{RQ4}. 
\subsubsection{Temporal Information Construction}
\begin{table}
\caption{Variants of Temporal Information Construction}
\begin{tabular}{@{}lccccc@{}}
\toprule
Variant                  & Toys   & Baby   & Tools           & Music & ML100K \\ \midrule
\modelname & \textbf{0.3649} & \textbf{0.2235} & \textbf{0.3623}          & \textbf{0.5935} & \textbf{0.3118}\\
$\mathcal{U}$ w\slash o T                   &  0.0103     & 0.0138      &  0.0106           & 0.0112 &0.1555 \\
$\mathcal{I}$ w\slash o T                   &  {\ul 0.1013}     & \text{\ul 0.0961}       &   \text{\ul 0.0836}           &\text{\ul 0.2785} & \text{\ul 0.2336} \\
\bottomrule
\end{tabular}%
\label{tab:without_time}
\end{table}
We develop two variants by dismissing the time vector in either Eq.~(\ref{eq:query_information}) or Eq.~(\ref{eq:neighbor_information}), i.e., users without time vectors or items without time vectors. The results are presented in Table~\ref{tab:without_time}. The observations are two-fold. Firstly, the performance of items without time is better than users without time. It implies that the temporal inference of user embeddings are rather important, which matches the intuition that the preference of users are dynamic while items are relatively more static.  Secondly, the performance deteriorates significantly in both variants, indicating again \modelname is able to model temporal effects of collaborative signals while also encoding sequences.

\subsubsection{Temporal Attention Weights Visualization}
\begin{figure}
    \centering
    \includegraphics[width=0.35\textwidth]{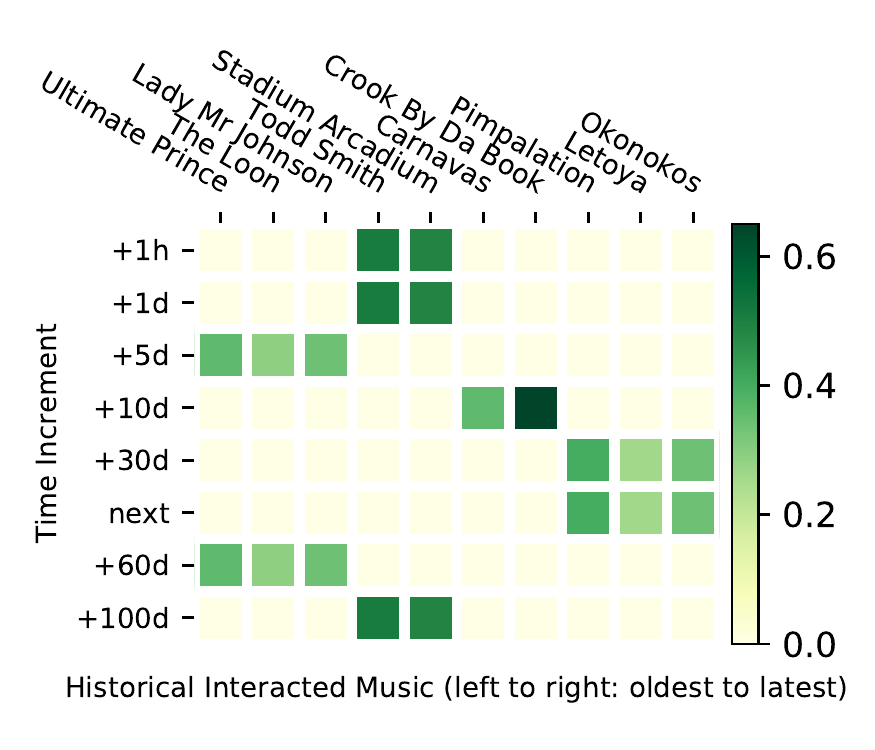}
    \vspace{-5mm}
    \caption{Temporal Attention Weights Visualization}
    \label{fig:att_vis}
\end{figure}
We visualize the attention weights of \modelname on the Music dataset for a user, which is shown in Figure~\ref{fig:att_vis}. Each row is associated with an increment (`h' for hour and `d' for day) from the last interactive timestamp $T=1159142400$, where `next' denotes the timestamp ($T$+34d) for the test interaction. Each column is associated with an item. We can observe that the attention weights for items are dynamic at different timestamps, which indicates the temporal inference characteristics of \modelname. Moreover, the time increments can be arbitrary values, which verifies its continuity. 
\subsubsection{Recommendation Results.} Besides the attention visualization, we also present a part of the recommendation results of the same user in Table~\ref{tab:rec_list_diff}. Additionally, we also show the results of SASRec and TiSASRec, which only leverage sequential patterns. We find that only \modelname can predict the ground truth item (\textit{Killing Joke}) in top-4 predictions at the time of interests. When time~(e.g., $T$+30d) becomes close to the predicting timestampe `next'~(i.e., $T$+34d), the ground truth item appears in the top-4 predictions.
We can observe that the top predicted items from SASRec are also recommended by \modelname, though in lower ranks. It again proves that \modelname can unify sequential patterns and temporal collaborative signals.
\begin{table}[]
\caption{Recommendation w.r.t. time increments after the last interaction at timestamp $T=1159142400$. `next' is the timestamp of the test interaction. The ground truth item is in red color. Items also predicted by SASRec and TiSASRec are in blue color.}
\label{tab:rec_list_diff}
\resizebox{0.47\textwidth}{!}{%
\begin{tabular}{lllll}
\toprule
Time & {Rank-1} & {Rank-2}               & {Rank-3} & {Rank-4} \\ \midrule
T+5d      & Letoya                &  {H. of Blue L.}      & Ult. Prince      & Veneer           \\
T+30d & {J. of A Gemini}   & Living Lgds.                      & {\color[HTML]{FE0000} Killing Joke} & \textcolor{blue}{Crane Wife}            \\
next & Buf. S.F.   & {\color[HTML]{FE0000} Killing Joke} & \textcolor{blue}{Empire}                              & \textcolor{blue}{Stadium Arc.}          \\
T+60d     & D. of Future P. & Even Now                 & L. Mks. Wd. & Przts. Author
\\ \midrule
SAS.   & \text{Crane Wife}        & \text{Empire}                   & H. Fna. Are    & You in Rev. \\
TiSAS. & \text{Crane Wife}        & \text{Empire}                   & WTE. P. S.  &  {Stadium Arc.} 
\\ \bottomrule
\end{tabular}%
}
\end{table}

\section{Conclusion}
In this paper, we design a new SR model, \modelname, to unify sequential patterns and temporal collaborative signals. \modelname is defined upon the proposed CTBG.
We apply a temporal kernel to map continuous timestamps on edges to vectors. Then, we introduce the \layername layer, which can infer temporal embeddings of nodes. 
It samples neighbors and learns attention weights to aggregate both node embeddings and time vectors. 
In this way, a \layername layer is able to encode both sequential patterns and collaborative signals, as well as reveal temporal effects. 
Extensive experiments on five real-world datasets demonstrate the effectiveness of \modelname.
\modelname significantly outperforms existing transformer-based sequential recommendation models.
Moreover, the ablation study and detailed analyses verify the efficacy of those components in \modelname.
In conclusion, \modelname is a better framework to solve the SR problem with temporal information. 

\section{Acknowledgments}
This work is supported in part by NSF under grants III-1763325, III-1909323,  III-2106758, and SaTC-1930941. This work is also partially supported by NSF through grant IIS-1763365 and by UC Davis. This work is also funded in part by the National Natural Science Foundation of China Projects  No. U1936213


\bibliographystyle{ACM-Reference-Format}
\balance
\bibliography{ref}


\end{document}